# PREPARATION AND CHARACTERIZATION OF PZT/PVDF COMPOSITES FILMS FABRICATED BY ELECTROSPINNING METHOD

Do Phuong Anh[1,*], Do Viet On[1], Dao Anh Quang[2], Nguyen Van Thinh[3], Nguyen Thi Anh Tuyet[4], Vo Thanh Tung[1] and Truong Van Chuong[1]

[1]University of Sciences, Hue University, Vietnam

[2]Institute of Reasearch and Development, Duy Tan University, Danang, Vietnam

[3]University of Technology and Education, University of Danang, Vietnam

[4]Faculty of Natural Sciences, Quy Nhon University, Binhdinh, Vietnam

E-mail: dpasophys@gmail.com; vttung@hueuni.edu.vn

**Abstract**

In this paper, we present some properties of Lead Zirconate Titanate/Poly(Vinylidene Fluoride) (PZT/PVDF) composites. The obtained results indicate that 0.9÷1.8 μm PZT/PVDF fibers involved films were achieved using electrospinning method. In other hand, the effect of grain size, content and other factors under the purview of Young's modulus and ferroelectric properties.

**Keywords:** PVDF, PZT, composites, electrospinning, ferroelectric, piezoelectric, nanofiber.

## 1. Introduction

The use of ceramic nanoparticles in various organic polymers may provide performance and multifunctional materials with applications in many technological fields such as actuators, sensors, solar cells, optoelectronic devices, etc.

Polyvinylidene fluoride (PVDF) is a fluorinated polymer with excellent mechanical and electric properties, which it was chosen as matrix due to their applications in a wide range of industrial fields. Research in the field of material science and engineering has been devoted to the development of innovative materials known as ceramic-polymer composites [2, 3, 8, 9, 13, 15, 16, 19-23, 25, 38]

Poly(vinylidene fluoride) (PVDF) is one of the semicrystalline polymers with at least five crystalline phases, its can form a different crystal depending on the condition of the crystallization. The different crystal structures are nonpolar α-phase, polar β -, γ -, δ- phase and ε – phase [4, 9, 12, 14-16, 18, 21, 25, 27-30, 35, 37, 39 ]. The β - phase crystal has all trans conformation that results in the most polar phase among other crystals, being used extensively in piezoelectric, pyroelectric and ferroelectric applications, because its piezoelectric activity is based on the dipole orientation within the crystalline phase.

A variety of experimental techniques have been developed to induce β-phase formation in PVDF. For example, Matsushige and Takemura showed that crystallization from the melt at pressures which exceed 350 MPa led to the formation of the β-form of PVDF. Uniaxial or biaxial drawing of PVDF films has also been shown to induce an α-β transition. In addition, a number of reports also indicate that nanoclays, carbon nanotubes, graphene oxides and metal oxides can induce the β-crystal formation in PVDF nanocomposites prepared via melt processing or solution processing. Besides, electrospinning also produces pre-dominantly the β-phase.

Ceramic-polymer composites are studied regarding possible preparation procedures, their ferroelectric properties and their tensile behavior. Furthermore, different arrangements of functional ingredients are investigated such as 1-3, 2-2 and 0-3 composites [1-6, 11, 18-21, 26, 29, 31- 34, 38, 40].

Especially composites with ceramic fillers randomly dispersed in a polymer matrix are broadly studied because of good preparation opportunities in order to vary the filler content, the filler and the matrix materials. Such composites are called 0-3 composites according to Newnham's connectivity notation. Then, the PZT ceramic particles with excellent piezoelectric properties were evenly dispersed in a three-dimensional connected flexible polymer matrix. This type of composite has excellent comprehensive properties

The influence of preparation processes on the structure and electrical properties of piezoelectric lead zirconate titanate/poly(vinylidene fluoride) composites was investigated. Electrospinning method can enhance obviously the ferroelectric, dielectric and piezoelectric properties of PZT/PVDF composites by reducing the void in piezoelectric composites and decreasing the interface defect between ceramic phase and polymer phase [4, 5, 8, 10, 36, 40].

## 2. Experimental section

*2.1 Materials*

The PVDF used was Aldrich chemistry (France), average Mw~534,000. The solvents used were N,N-dimethylformamide (DMF, Merck 99.5%) and acetone (Merck, 99.7%). The PZT ceramic ($d_{33}$ = 317 pC/N, $\varepsilon_r$ = 1800), which was prepared via the conventional solid-state reaction method, PZT powers are prepared with a mortar and pestle, the mean particle size of the PZT powders is about 200 – 500 nm (Figure 1).

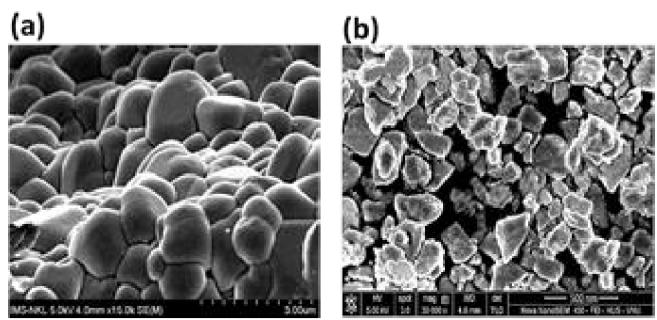

**Figure 1**. SEM images of particle size of the PZT before and after grinding

*2.2 Devices*

The crystalline structure analysis was performed at room temperature using an X-ray diffraction system (XRD, Bruker D8 Advance, Germany) with Cu K$_\alpha$ radiation (λ= 0.154 nm) and Fourier transform infrared (FTIR, Prestige 21 Shimadzu, Japan, Department of Chemical, Hue University of Education) spectroscopy, respectively. The surface morphology was examined using a scanning electron microscopy (SEM, Nova NanoSEM 450-FEI) operated at an accelerating voltage of 10–20 kV), the Raman spectra was recorded by a Micro Raman Spectrophotometer (JASCO Raman NRS-3000) using 633 nm excited laser at room temperature. The polarization-electric field (P−E) hysteresis loops were measured with a HIOKI 3532 (Radiant Technologies) which testing unit connected with a high-voltage interface. The electrospinning apparatus was assembled by Hue University of Science (Vietnam); it consisted of a high-voltage power supply, an automated syringe pump (KDS101, KD Scientific, USA) with speed from 1mL/h to 999 mL/h and a 10-mL syringe with a 25-gauge diameter of needle (~0.5 mm); The high electric potential ($E_G = 0 \div 30$ kV) and the stretching force ($F_G$) applied during electrospinning, a steel rotating (E-HUSC-01, Hue University of Science, Vietnam) with 1500 rpm mandrel was used as the collector

*2.3. Process*

*Preparation of the PVDF sol solution:* The PVDF precursor solutions 16 wt% were prepared by dissolving the PVDF powder in 3:1 DMF/acetone solution. The solution was heated at 65°C in silicon oil bath for 30 min to completely dissolve the polymer.

**Table 1.** Samples of PZT/PVDF

| Samples No. | Name of products | Composite compound ratio | $\rho$ (g/cm$^3$) |
|---|---|---|---|
| | F16 | PVDF 16 wt% + 0wt% PZT | 1.7 |
| | F-P5 | PVDF 16 wt% + 5wt% PZT | 1.8 |
| | F-P10 | PVDF 16 wt% + 10wt% PZT | 2.1 |
| | F-P15 | PVDF 16 wt% + 15wt% PZT | 2.4 |
| | F-P20 | PVDF 16 wt% + 20wt% PZT | 2.8 |
| | F-P25 | PVDF 16 wt% + 25wt% PZT | 3.1 |

*Fabrication of the PZT/PVDF composite films by electrospinning:* The PZT, which was added to the PVDF solutions at concentration of 5, 10, 15, 20, and 25wt%, respectively (Table 1). Solubilization was carried out at 65°C under ultrasound for 1 hour.

The solution flow rate was precisely maintained at 1.5 mL/h using the syringe pump. A 12-cm distance and an applied voltage of 14 kV were maintained between the needle and the collector [19].

The PZT/PVDF thin films were deposited by spin coating the solution after 4 h on the silver substrates at room temperature. The wet films were annealed at 135 °C for different durations in an oven.

Besides on, these composite films was found that the number of particles increases with the increase in the PZT volume fraction. There is a non-uniformity observed in particle size of the samples due to the presence of different particle sizes of PZT. The SEM micrographs also reveals that there is an increase in the reinforcement of PZT particles in the PVDF matrix which indicates that there is a change in volume fraction, Fig 2 [18].

# 3. Results and discussion

## 3.1 Crystalline phase analysis

### 3.1.1 Surface morphology

Figure 2a show the SEM micrographs of bare PVDF and PVDF with different wt% PZT. Smooth and bead-free solid fibres were obtained for all the samples. In order to estimate fibre size, the nanostructural photos of the composites were analyzed using Lince software. The fibre size located between 0.45 µm and 1.8 µm and gathered round the top of Gauss fitting plot.

The electrospinning parameters can be adjusted to control final PVDF fiber characteristics. One of the important parameters to consider is molecular weight of PZT. Figure 2 (a-m) shows the SEM micrographs and fiber diameter distribution of the various PVDF fibers at different PZT contents. During this sintering process, PZT/PVDF nanofibers become less flexible. At a higher voltage, the jet draws away more solution, which requires a higher flow rate to create a stable Taylor cone. Too low of a flow rate creates a break in the jet as more solution collects on the tip causes excess solution to fall to the ground.

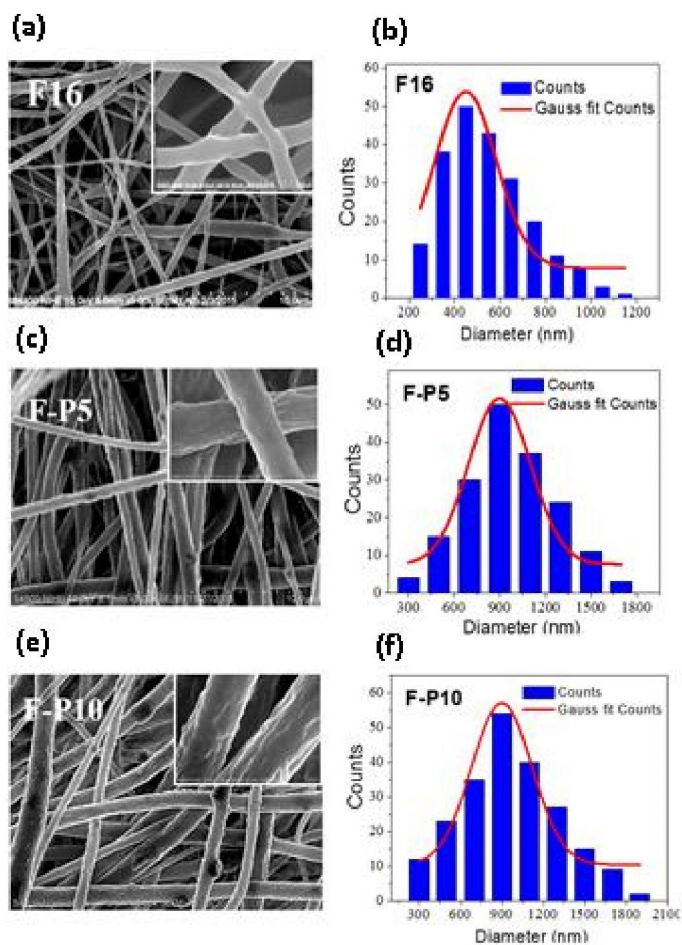

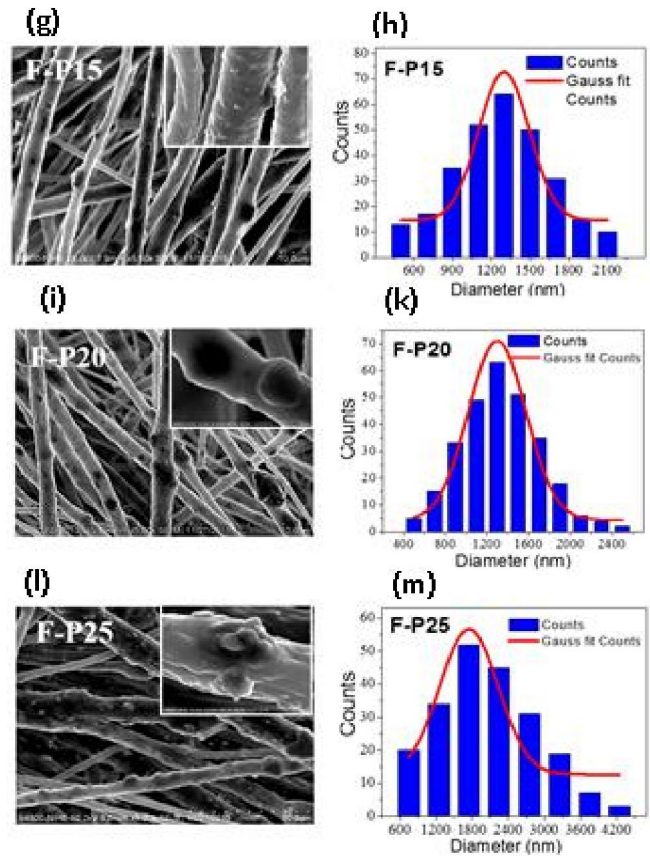

**Figure 2**. SEM images and grain size distribution for PZT/PVDF composites with different PZT: (a-b) 0 wt%, (c-d) 5 wt%, (e-f) 10 wt%, (g-h) 15 wt%, (i-k) 20 wt%, (l-m) 25 wt%.

Figure 2 shows, these structures were morphologically analyzed by scanning electron microscopy (SEM) with 900÷1800 nm diameter of PZT/PVDF, these process were achieved by maximum PZT the concentration (sign: F-P25), the spraying technique is not possible with higher PZT concentrations of 25wt%.

The effect of the PZT content on the morphology and average diameters of the PZT/PVDF nanofibers is shown in Fig. 2. To further understand the impact of PZT on PVDF, it had an average diameter of 450 nm (see in Figure 5(a-b)). However, an increase in the PZT content resulted in the formation of the gradual diameter of nanofibers increases. The diameter of nanofibers was significantly larger than mean difficult in spray coating on thin film [11, 22]. The PZT doping will significantly benefits the β-phase crystallization of PVDF composites and remnant polarization, however, if the PZT-doping concentration is too high, will damage the film quality, increase the thickness of film spin-coating, make the PZT/PVDF film break up [2]. A complete film can be achieved with 25 wt% PZT contents. On other hand, PZT can effectively improve the piezoelectric properties of PVDF.

*3.1.2 FTIR, XRAY and RAMAN, Crystalline structure*

Figure 3a shows the FTIR spectra of aligned electrospun PVDF fibers with different PZT contents (0–25 wt%); here the enhancement of β phase (infrared band at 478, 510, 840, 1072, 1274, and 1404 $cm^{-1}$) with increasing PZT content can be observed [6].

The FTIR spectrums of the nanofilms (Figure 3a), the absorption peaks at 1274, 1072 and 840 cm$^{-1}$ are typical vibration characteristics of crystalline β phase. It is notable that doping PZT into PVDF has a distinctive effect of both enhancing the crystalline β phase and weakening the crystalline α phase. This effect is closely related to the concentration of PZT. For the spectrum of higher PZT concentrations doped PVDF, the crystalline β phase absorption bands at 1274, 1072 and 840 cm$^{-1}$ improve compared with lower concentrations [1, 6, 9-14, 20, 22, 24-31, 37, 40].

In order to confirm the crystalline phase of the samples and the influence of PZT ceramic particles in the polymer matrix an X-ray diffraction analysis was carried out in the range 10° to 70° as shown in Fig. 3b. The PZT powder sample exhibits characteristic peaks observed at 2θ = 21.3°, 31.1°, 38.4°, 44.5°, 50.2°, 55.3°, and 64.7° corresponding to the planes with Miller indices of (001), (110), (111), (200), (201), (211), (022). For PVDF films, the shoulder peak at the 2θ value of about 17.2° corresponds to (020) and (110) planes of crystalline α phase and peak at 20.4° indicates the PVDF β-phase [1, 5, 6, 7, 11, 12, 16, 17, 20-31, 40].

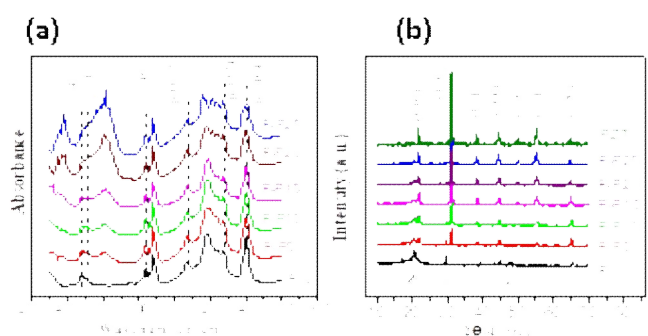

*Figure 3. FTIR spectra and XRD curves of the PZT/PVDF composite film with different wt% PZT*

By comparing the PZT powder with of the PZT-PVDF composites, it is found that there is no additional evidence of structural change, XRD analysis is in good agreement with FTIR analysis.

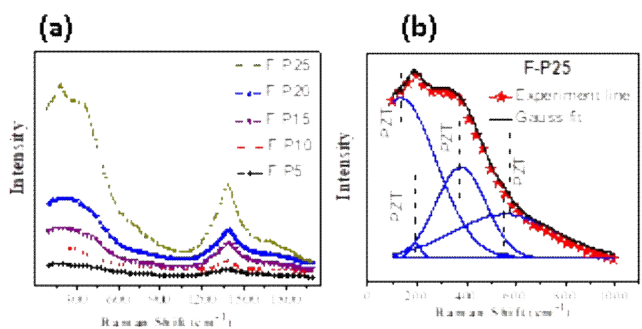

*Figure 4. Raman spectrum of PVDF 16%wt with different contant of PZT*

We prepared the 0-3 composites of pure PVDF and doped them with PZT the concentration (5-25 wt%). The PZT content was done to increase the conductivity of the matrix as this increased conductivity will enhance the poling of ceramic at a later stage.

Figure 4 shows that the Raman spectra of the PZT-PVDF film are completely dominated by the PZT material. The peaks with wave numbers of 130, 185, 374, 545 and 573 cm$^{-1}$ were assigned to the tetragonal PZT phase. These peaks are identified as A1(1TO), E(2TO), A1(2TO), E(3TO), and A1(3TO) modes, respectively [6, 7, 12, 25, 26, 30, 40]. It has also been shown that when the PZT increases several times, it is shown that the peak intensities are from 600 cm$^{-1}$ to 1000 cm$^{-1}$, most of the PZT crystals have been trapped in the PVDF matrix during crystallization. This is clearly seen through the SEM image of the composite.

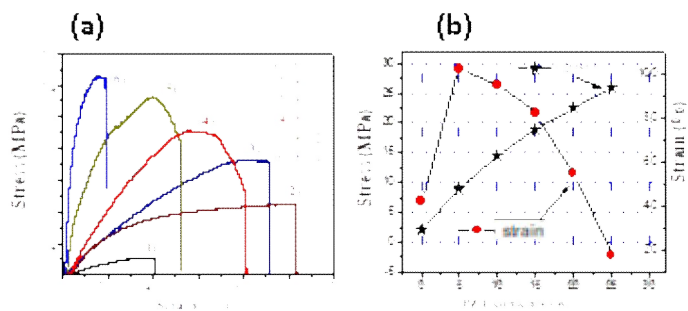

*Figure 5. Tensile stress-strain properties of PZT/PVDF nanofiber composite films prepared with different levels of PZT content. The inset shows the results of Young's modulus for PZT/PVDF nanofiber composite films.*

Figure 5 shows the typical stress/strain curves of all samples. The Young's moduli of the PZT/PVDF composite nanofibres were much higher than that of the bare PVDF nanofibre, especially for the 15 wt% samples. A slight decline in mechanical properties for the 25 wt% sample may have been due to its low crystallinity, as confirmed by the XRD and Raman spectrum results.

PZT doping will significantly benefits the β-phase crystallization of PVDF nanocomposites and remnant polarization, however, if the PZT-doping concentration is too high, will damage the film quality, increase the thickness of film spin-coating, make the film less transparent, warp the substrate, and in the worst case, make the PZT/PVDF film peel off. This may have been due to the addition of PZT hindering the movement of the PVDF molecular chain, which reduced the elasticity and the elongation of the composite [9, 12, 17, 24, 39].

PZT can effectively improve the piezoelectric properties of PVDF, although excessive PZT doping will limit the flexibility of composite film as it sustains less elongation just before failure. However, it had been found that high doping concentrations will increased remnant polarization because there will be a significant increase in the dielectric constant and thus the piezoelectric signal will increase [12, 23].

*3.2 Ferroelectric properties of the PZT/PVDF films*

A series of P–E hysteresis loops for PZT/PVDF composite measured at room temperature were shown in Fig. 6. The results showed that the composites with PZT volume fraction 5 to 25 wt% exhibited non-saturated P–E hysteresis loop.

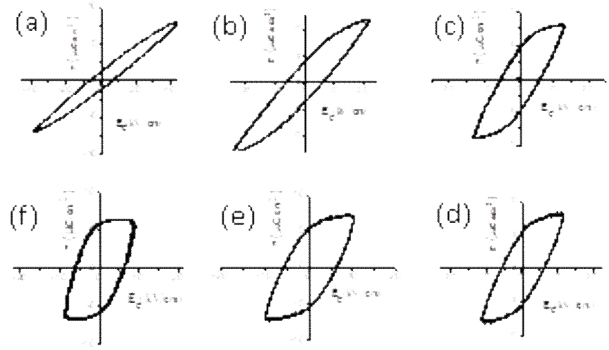

**Figure 6.** The P−E hysteresis loops for F16(a) and F16-x%PZT films with different concentrations of PZT, x= 5 (b), 10 (c), 15 (d), 20 (e), 25 (f) wt%

It is also be noted that when increasing the PZT content to beyond, values of remnant polarization increases and coercive field decreases to a finite value as shown in table 2. The results thus suggested that distribution of PZT granules in PVDF matrix played a significant role in controlling of ferroelectric behavior. The ferroelectric properties of the PZT/PVDF films with different concentrations of PZT between 5 and 25 wt% were investigated.

Table 2. Ferroelectric properties of the PZT/GO/PVDF films

| Samples No. | Name of products | Composite compound ratio | Remnant Polariation ($P_r$) µC/cm$^2$ | Coercive field ($E_C$) kV/cm |
|---|---|---|---|---|
| | F16 | 0 | 2.1 | 9.3 |
| | F-P5 | 5 | 4.3 | 7.2 |
| | F-P10 | 10 | 4.8 | 6.1 |
| | F-P15 | 15 | 5.3 | 6.1 |
| | F-P20 | 20 | 5.9 | 5.7 |
| | F-P25 | 25 | 6.2 | 4.8 |

The gradual increase of $P_r$ is caused by the crystallinity enhancement of polar β-phase with increasing of doped PZT and annealed techniques. The correlation between the amorphous phase contents and ferroelectric properties of PZT thin films was further discussed from the viewpoint of local chemical structures, Fig.6 [7, 17-21, 34, 35, 40].

A remnant polarization increasing from 2,1 to 6.2 µC/m$^2$ were achieved by optimizing the concentration of 25 wt% of PZT, while simultaneously reducing the coercive field at the same time. The modification of the physical properties can be explained by alterations in PVDF fiber due to drawing and poling and to the presence of PZT (Table 2), [10, 36].

## 4. Conclusion

In this study, PZT-PVDF composites with 0–3 connectivity was successfully fabricated from a series of PZT volume fractions from 5 to 25 by electrospinning method. With the increase of the PZT content, an enhancement of the β phase was observed. The PZT-PVDF composites open new opportunities to develop a large scale manufacturing of flexible devices, piezoelectric devices, sensors, force transducer, and so on with technological applications.

Experimental density values of the composites were found to be harmonized with calculated density values. SEM observations revealed a homogeneous mixture of PZT-PVDF phases. The values of stress - strain coefficient and ferroelectric character obtained in this work revealed that it may be more suitable for widely future applications.

**Acknowledgements**